\documentclass[aps,prb,twocolumn,showpacs,preprintnumbers,amsmath,amssymb,superscriptaddress]{revtex4-1}
\usepackage[dvips]{graphicx}% Include figure files
\usepackage{dcolumn}% Align table columns on decimal point
\usepackage{bm}% bold math
\usepackage{color}
\usepackage{amsmath}
\usepackage{amssymb}
\usepackage{amsbsy}
\usepackage[utf8]{inputenc}

\usepackage{ifpdf}

\ifpdf
\usepackage[pdftex]{hyperref}
\else
\usepackage[hypertex,ps2pdf,dvips,colorlinks,urlcolor=blue,citecolor=blue]{hyperref}
\fi

\newcommand{\be}{\begin{equation}}
\newcommand{\ee}{\end{equation}}
\newcommand{\bea}{\begin{eqnarray}}
\newcommand{\eea}{\end{eqnarray}}

\def \ket #1{\left| #1 \right\rangle}

\def \scalarprod #1#2{\left \langle #1 \right. \left| #2 \right\rangle}

\def\bes{\begin{subequations}}
\def\esu{\end{subequations}}
\def\erf{\eqref}

\begin{document}

\frenchspacing
 
\title{Spin fluctuations after quantum quenches in the $S=1$ Haldane chain: numerical validation of the semi-semiclassical theory}

\author{Mikl\'os Antal Werner}
\affiliation{BME-MTA Exotic Quantum Phase Group, Institute of Physics, Budapest University of Technology and Economics, H-1111 Budapest, Hungary}

\author{C\u at\u alin Pa\c scu Moca}
\affiliation{BME-MTA Exotic Quantum Phase Group, Institute of Physics, Budapest University of Technology and Economics,
H-1111 Budapest, Hungary}
\affiliation{Department of Physics, University of Oradea, 410087, Oradea, Romania}

\author{\"Ors Legeza} 
\affiliation{Strongly Correlated Systems Lend\"ulet Research Group, Institute for Solid State Physics and Optics,MTA Wigner Research Centre for Physics, P.O. Box 49, H-1525 Budapest, Hungary}

\author{M\'arton Kormos}
\affiliation{BME-MTA Statistical Field Theory Research Group, Institute of Physics, Budapest University of Technology and Economics,
H-1111 Budapest, Hungary}

\author{Gergely Zar\' and}
\affiliation{BME-MTA Exotic Quantum Phase Group, Institute of Physics, Budapest University of Technology and Economics, H-1111 Budapest, Hungary}

\date{\today}

\begin{abstract}
{
We study quantum quenches in the $S=1$ Heisenberg spin chain and show that the dynamics can be described by the recently developed semi-semiclassical method based on particles propagating along classical trajectories but scattering quantum mechanically. We analyze the non-equilibrium time evolution of the distribution of the total spin in half of the system and compare the predictions of the semi-semiclassical theory with those of a non-Abelian time evolving block decimation (TEBD) algorithm which exploits the SU(2) symmetry. We show that while the standard semiclassical approach using the universal low energy scattering matrix cannot describe the dynamics, the hybrid semiclassical method based on the full scattering matrix gives excellent agreement with the first principles TEBD simulation.
}
\end{abstract}

%\pacs{00.00.xx}

\maketitle

\section{Introduction}
Understanding non-equilibrium dynamics in interacting quantum many-body systems is one of the major challenges in today's statistical physics \cite{Cazalilla2010,Polkovnikov2011,Eisert2015,Calabrese2016}.
The formation of current-carrying steady states
\cite{Prosen2015,Bertini2016,Bernard2016a,Collura2018}, the details of the thermalization process \cite{Moeckel2008,Bertini2015a}, entropy production \cite{Calabrese2005,Schuch2008,Alba2017,Nahum2017,Collura2018a}
or the interplay with disorder \cite{Znidaric2008,Bardarson2012,Nanduri2014,Serbyn2014,Vasseur2016} and topology \cite{Halasz2012,Mazza2014,DAlessio2015,McGinley2018} are just a few examples of the intriguing open questions which, due to recent breakthroughs in quantum simulation, can now be investigated experimentally.

While experimental results are abounding \cite{Kinoshita2006,Trotzky2012,Gring2012,Chenau2012,Langen2015,Kaufman2016}, available theoretical tools are quite limited in number and power.
One dimensional systems represent in this regard an exception and a theoretical testing ground for all methods and investigations, especially since    
in one dimension powerful analytical and numerical methods such as Bethe Ansatz\cite{Korepin_book},  bosonization \cite{Haldane_Bosonization, Giamarchi_book}, Density Matrix Renormalization Group (DMRG) \cite{White1992, Schollwock2011} exist to study equilibrium properties in detail. These methods can be extended to non-equilibrium steady states\cite{Prosen2009, Cui2015}, while dynamics under non-equilibrium conditions 
can be efficiently simulated by the tDMRG \cite{White2004} and the Time-Evolving Block Decimation (TEBD) algorithms \cite{Vidal2007}. However, in dynamical simulations “exactness” or precision is typically lost after relatively short
times due to the rapidly increasing entanglement of the state.

Recently, a semi-semiclassical approach (SSA) has been proposed to study generic, gapped one dimensional quantum systems at \emph{longer} times, and demonstrated on the sine--Gordon model \cite{PascuMarciGergo2017}. As a generalization of the original semiclassical approach of 
Sachdev, Young, and Damle \cite{SachdevYoung1997,SachdevDamle1997}, the semi-semiclassical method treats trajectories of quasiparticles classically, but it accounts for the precise quantum evolution of the  internal degrees of freedom fully quantum-mechanically, and allows one to capture the associated quantum 
entanglement generation as well as simultaneous probabilistic processes. Though approximate, the SSA method is versatile, conceptually simple, and has also been extended to study dynamics in non-equilibrium steady states (NESS) within the nonlinear sigma model \cite{Marci_Pascu_Gergo_NESS}.

\begin{figure}[b]
 \includegraphics[width = 0.45 \textwidth]{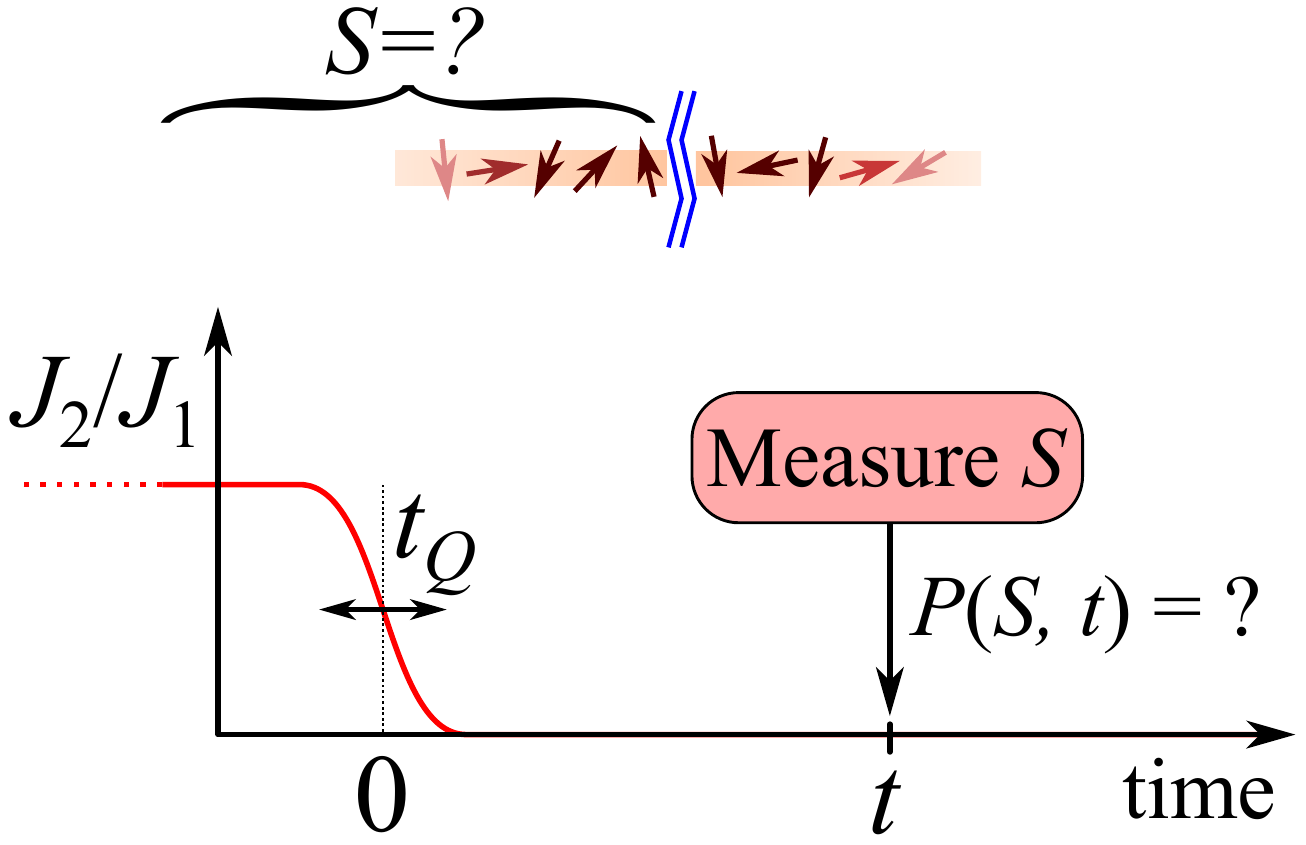}
 \caption{
 \textit{(Color online)} Visualization of the quench protocol. The biquadratic coupling $J_2$ is turned off at $t = 0$ within a switching time $t_Q.$ Later, at time $t,$ 
 the system is cut in two parts and the total spin $S$ of the left part is measured. The statistics of the measurement is described by the spin distribution $P(S,t)$. }\label{fig:quench_protocol}
\end{figure}

Here we investigate quasiparticle creation and spin propagation after a quantum quench within the Haldane-gapped phase of the the antiferromagnetic spin-1 Heisenberg chain
and compare the predictions of the semi-semiclassical method with TEBD  simulations in detail. In particular, 
we consider the  $S=1$ Haldane chain with a time-dependent biquadratic interaction $J_2(t)$
\begin{equation}\label{eq:Hamiltonian}
 \hat{H}(t) = \sum_{i} J_1 \; \vec{S}_i \cdot \vec{S}_{i+1} + J_2(t) \left( \vec{S}_i \cdot \vec{S}_{i+1} \right)^2 \;.
\end{equation}
Here $J_2(t) = J_2 \gamma(t)$, and the function $\gamma(t)$ changes from $1$ to $0$ around $t=0$. 
We start from the ground state of $H(-\infty)$ and perform infinite volume TEBD (iTEBD) simulations~\cite{Vidal2007} 
to obtain the full time dependent wave function of the chain, $|\psi(t)\rangle $. To reach sufficiently long times, 
comparable with the quasiparticle collision times,  we need to  exploit non-Abelian symmetries in our simulations. 

The quench protocol described above generates a gas of (entangled pairs of) quasiparticles in the final state.   
Cutting the infinite chain into two at time $t$ and then measuring the spin distribution on one side, $P(S,t)$, 
we can  explore  the propagation and collision of these quasiparticles  (Fig.~\ref{fig:quench_protocol}). 
We can, in particular, compute $P(S,t)$ in terms of the semi-semiclassical approach, and compare it to the 
results of TEBD simulations to find an astonishing agreement. 
As we shall also demonstrate, a theory based on completely reflective quasiparticle collisions \cite{SachdevDamle1997, Damle2005,RappZarand,Evangelisti2013,Kormos_Zarand_PRE_2016} 
is not able to account for the observed behavior, and the full semi-semiclassical approach  is needed to get agreement 
with TEBD computations. 

The paper is structured as follows. In Sec. \ref{sec:Basic_concepts} we overview the basics of the microscopic non-Abelian iTEBD simulations and the SSA method. In Sec. \ref{sec:Short_times} the short time ballistic behavior after the quench is analyzed. Then in Sec. \ref{sec:Long_times} we discuss the collision dominated regime of the dynamics. In Sec. \ref{sec:pert} we develop a perturbative quench theory and test its scope of validity. Our conclusions are summarized in Sec. \ref{sec:summary}.

%For a sudden quench,  $\gamma(t)$  is just a $\Theta$ function, while for smooth quantum quenches it 
%varies within the scale of the quench time, $t_Q$. 

\section{Basic concepts and numerical methods}\label{sec:Basic_concepts}

Before presenting the main results, let us shortly review the two methods used to investigate the quench dynamics. 

\subsection{Non-abelian TEBD lattice simulations}

Here we discuss our non-abelian  TEBD algorithm only briefly, since our flexible and general way of treating non-Abelian symmetries in MPS simulations will be presented in a separate publication \cite{Werner_future}.

The post-quench dynamics of the system is described by the microscopic many-body Schr\"odinger equation,  
\begin{equation}\label{eq:Schroedinger}
 i \partial_t \ket{\Psi(t)} = \hat{H}(t) \ket{\Psi(t)}
\end{equation}
which we  solve numerically by means of the infinite chain time-evolving block decimation (iTEBD) algorithm \cite{Vidal2007}.
The TEBD algorithm describes the real time dynamics of the system based on the matrix product state (MPS) description of the quantum state $\ket{\Psi(t)}$ \cite{Schollwock2011, Orus2014},
%and represents $|\psi(t)\rangle$ as a  product, 
\begin{equation}
|\Psi\rangle =  \sum_{ \{ \sigma_l \} }   (\dots  M^{[l-1]}_{\sigma_{l-1}} M^{[l]}_{\sigma_l}\dots) \; |    \{ \sigma_l\} \rangle \;,
\label{eq:MPS}
\end{equation}
where the $|\sigma_l\rangle$ with $\sigma_l = \lbrace 0,+,- \rbrace$ refer to the three quantum states of spin $S_l$ at site $l$.

\begin{figure}
 \includegraphics[width = 0.45 \textwidth]{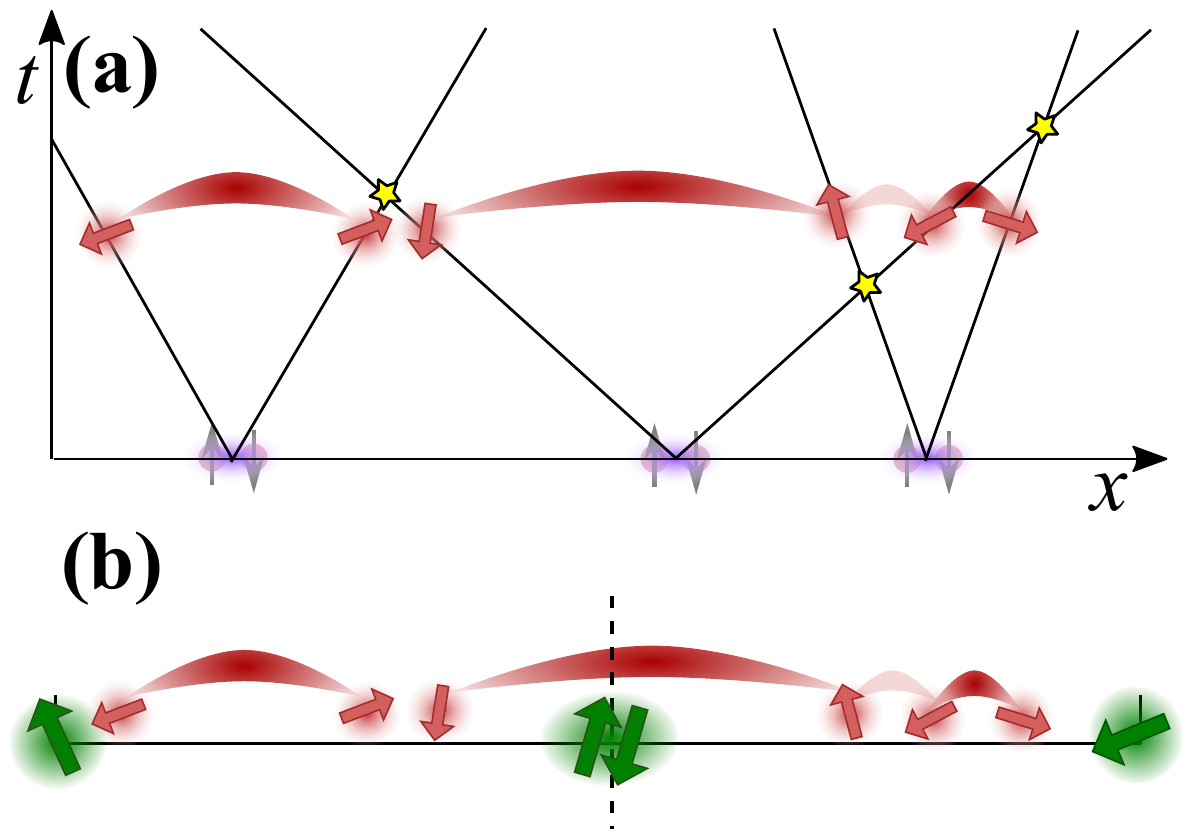}
 \caption{\textit{(Color online)} The semiclassical quasiparticle picture. \textbf{(a)} World lines of the excited quasiparticle pairs. Entanglement structure of the gas is indicated by arched red stripes. 
 At $t=0$ singlet pairs with zero total momentum are excited. After the quench the quasiparticles collide with each other, the scattering events are denoted by stars. 
 \textbf{(b)} Visualization of the spin structure if the system is cut in two parts. The topological edge-spins are denoted by large green arrows, while the quasiparticle spins are denoted by 
 small red arrows. At the cut, the two ``virtual'' edge-spins form a singlet. The total spin of the left part contains contributions of the edge-spins, the quasiparticles and vacuum fluctuations.  }\label{fig:sketch}
\end{figure}

The MPS factorization of the quantum state $\ket{\Psi}$ relies on Schmidt decomposition \cite{Schmidt1907}.
Let us  now focus on the case  where  $\ket{\Psi}$ is a spin-singlet,  and  cut the chain into two halves between sites $l$ and $l+1$, i.e.  treat the Hilbert space of the full chain as the product of its left and right halves. The Schmidt decomposition of  $\ket{\Psi}$ then reads as\cite{McCulloch,Singh_Vidal2010}
\begin{equation}\label{eq:Schmidt_decomp}
 \ket{\Psi} = \sum_{t_l} \frac{\Lambda^{[l]}_{t_l}}{\sqrt{2 J_{t_l} + 1}} \sum_{m_l=-J_{t_l}}^{J_{t_l}} \ket{t_l,m_l}_{\mathrm{left}} \ket{t_l,\overline{m}_l}_{\mathrm{right}} \, ,
\end{equation}
where $ \ket{t_l,m_l}_{\mathrm{left}} $ and $ \ket{t_l,\overline{m}_l}_{\mathrm{right}} $ are the so-called Schmidt-pairs \cite{footnote_right_state}. Here $t_l$ labels multiplets, while $m_l$ refers to  internal states within this  multiplet having a total spin $J_{t_l}$. The values $\Lambda^{[l]}_{t_l}/\sqrt{2 J_{t_l}+1}>0$ are the so called Schmidt values and are independent of the internal label $m_l$. 

\begin{figure}[t]
 \includegraphics[width = 0.45 \textwidth]{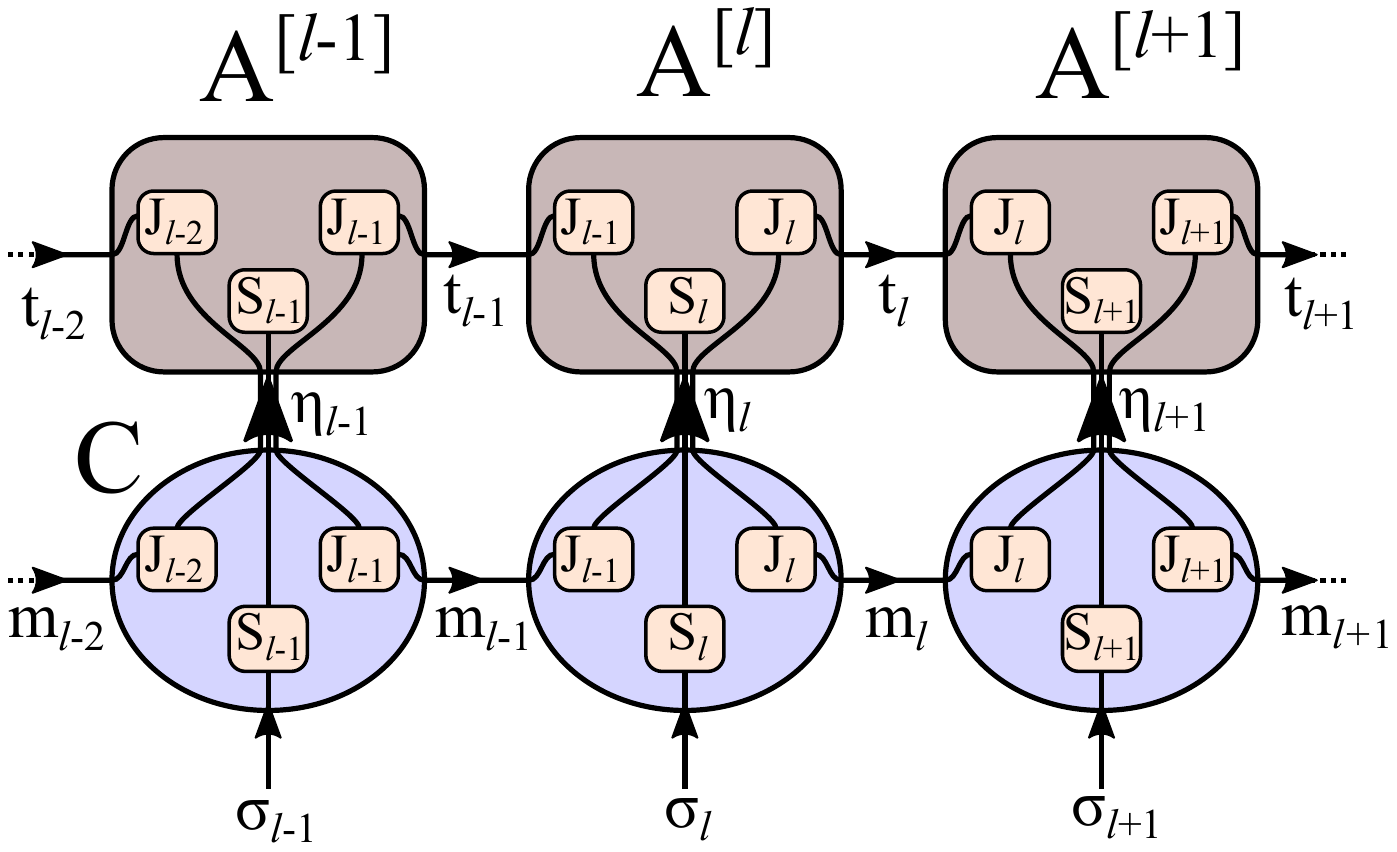}
 \caption{\textit{(Color online)} Two layer structure of the SU(2) invariant non-Abelian matrix product state. The lower layer contains the Clebsch--Gordan coefficients and is unaffected by time evolution. Representation indices are shown as bubbles within the tensors: these indices indicate the sparse block structure these tensors.}\label{fig:NAMPS}
\end{figure}

The non-Abelian MPS (NA-MPS) representation of $|\Psi\rangle $ can be constructed  by relating  neighboring left Schmidt states 
while moving  the cut position forward by one site, 
\begin{multline}
\label{eq:NA_MPS_rotation}
 \ket{t_l,m_l}_{\mathrm{left}} = \sum_{t_{l-1}, \eta_l} A^{[l] \, t_{l}}_{t_{l-1}\, \eta_{l}} \sum_{m_{l-1},\sigma_{l}} C^{m_l\, \eta_l}_{m_{l-1} \, \sigma_{l}} \\ \ket{t_{l-1},m_{l-1}}_{\mathrm{left}}  \ket{\sigma_{l}}\, .
\end{multline}
%where $\ket{\sigma_l}$ denote the three states of spin  $S_l = 1$ at site $l$. 
Here the tensor $C^{m_l\, \eta_l}_{m_{l-1} \, \sigma_{l}}$ contains the Clebsch--Gordan coefficients $\scalarprod{J_{l-1},m_{l-1};S_{l},\sigma_l}{J_{l},m_{l}}$, while the superindex  $\eta_{l}\to \{J_{l-1}, S_{l}, J_{l}\} $  
runs over  allowed  values of the three spins \cite{footnote_on_outer_multiplicity}.
An iterative  application of Eq.~\eqref{eq:NA_MPS_rotation}
 yields the non-Abelian MPS (NA-MPS) representation 
\eqref{eq:MPS} with 
\begin{equation}
(M^{[l]}_{\sigma_l})^{t_l,m_l}_{t_{l-1},m_{l-1}} =   
\sum_{ \eta_l} (A^{[l]})^{t_l}_{t_{l-1}\, \eta_l}  \,C^{\,m_l \, \eta_l}_{\,m_{l-1}\,\sigma_l} \,, 
\end{equation}
corresponding  to the two-layer NA-MPS structure sketched in Fig.~\ref{fig:NAMPS}. 
The tensors $A^{[l]}$ describe the transformation at the level of multiplets, and  the contraction of the index $\eta_{l}$  between 
$A^{[l]}$ and $C$ ensures that the values of the three representation indices in the two tensors match. 

The upper layer of the tensors $A^{[l]}$ contains all  important physical information on the state $|\Psi\rangle$, 
and our TEBD  time-evolver operators act only  on this  upper NA-MPS layer  (see Appendix  \ref{appendix:TEBD}).
In other words,  all expensive contractions within  the Clebsch--Gordan layer are eliminated. 

Here  we use the infinite-chain TEBD algorithm that applies for translation invariant states\cite{Vidal2007}. In this case, both the tensors  $A^{[l]}$ and the Schmidt values $\Lambda^{[l]}_{t_l}$ are independent of the site index $l$. Numerically, this one site translation invariance is, however,  lowered to a two-site translation invariance due to the Suzuki--Trotter time evolution scheme \cite{Trotter,Suzuki}, i.e. the tensors on the even and odd sublattices are slightly different.

The TEBD simulation provides   the time-dependent tensors $A(t)^{ \, t_{l}}_{t_{l-1}\, \eta_{l}}$ and the Schmidt values $\Lambda(t)_{t_l}$ (for details see Appendix \ref{appendix:TEBD}). The distribution of the total spin of the half chain is then simply related to the Schmidt values, 
\begin{equation}
 P(S,t) = \sum_{t_l\,| \, J_{t_l} = S} |\Lambda(t)_{t_l}|^2 \; .
\end{equation}
As shown in later sections, this distribution contains the  essential information on the spin structure and spin entanglement of excited quasiparticles in the post-quench state.

\subsection{The semi-semiclassical approach} 
%Though \eqref{eq:Schroedinger} gives in principle the full dynamics of the system, 
The  numerical resources required for   TEBD  grow exponentially in time, and make the microscopic simulations  tractable only for short times. However,  the single-site resolved knowledge of the quantum state is not necessary for answering many questions.
 The model \eqref{eq:Hamiltonian} is known to be gapped~\cite{Haldane1, AffleckWhite2008}, and its low-energy excitations 
 are  $S=1$ triplet quasiparticles \cite{AffleckWhite2008} described by the $O(3)$ nonlinear sigma model, an integrable relativistic field theory. Low energy quasiparticles  therefore have a (close to) relativistic dispersion  
\be
\label{eq:disprel}
\omega_{q} = \sqrt{\Delta^2 + c^2 q^2}\,,
\ee
where the gap and the ``speed of light" are given in terms of the microscopic parameters of the Hamiltonian as\cite{AffleckWhite2008}
\begin{equation}
\label{eq:gapc}
\Delta \approx 0.4105 \,J_1\, , \quad c \approx 2.472 \,J_1a\,.
\end{equation}
The finite gap ensures that, in the case of small quenches, the post-quench state is a dilute gas of quasiparticles.

 From the locality and the translation invariance of the Hamiltonian we can also conclude that shortly after the quench the state consists of spatially localized uncorrelated quasiparticle pairs. Moreover,  since the quench protocol conserves  SU(2), 
 these quasiparticle pairs must form SU(2) singlets. These assumptions form the basis of the semi-semiclassical approximation~\cite{PascuMarciGergo2017}, where the quasiparticles' spatial degrees of freedom are treated classically in a Monte Carlo sampling of the possible world-line configurations, while their  internal spin states are followed at the quantum level. 
  As sketched in Fig. \ref{fig:sketch}, the evolution of the quasiparticle spin states can be described by consecutive application of the two-particle S-matrix at collision events --  if the gas is dilute enough. 
  
 The condition for the applicability of semiclassical  approach is that the mean interparticle distance must  be larger than the Compton wavelength of the particles\cite{footnote_debroglie}
\be
\label{eq:lowdens}
\rho^{-1} >  c/\Delta\; .
\ee 
As we shall see, this condition is satisfied even for relatively large quenches in our model. 

This semi-semiclassical formalism allows us to compute the time dependent spin distribution of a half-chain,  $ P(S,t) $.
Half-chain spin fluctuations have two sources in the semi-semiclassical approach: (1)  cutting the vacuum state of an infinite Haldane chain into two  gives rise to a non-trivial spin distribution, $P_0(S_0),$ and (2) quench generated quasiparticles carry spins (and entanglement) across the cut.  
 
 To determine the first contribution, we first constructed the post-quench ground state with $J_2 = 0$ using TEBD, 
 cut it into two, and  determined $P_0(S_0)$ from the Schmidt values. 
 As shown in Table \ref{tab:P0}, we clearly observe the presence of two topologically protected 
 spin 1/2 end spins after the cut \cite{AKLT,Kennedy1990}: these yield   a triplet state  with almost 74\% probability and  a singlet state with more than  24\% likelihood. 
 %\red{In an ideal, very long chain, these probabilities would be just 75\% and 25\%. However, here the cut itself  also generates quasiparticles, although the probability of these processes less than about 2\%.}\WM{I think the statement is not phrased correctly. As long as the ground state has finite correlation length (not like in the AKLT point), and the exchange at the cut is not set to $0$, the vacuum-fluctuations change the $P_0(S_0)$ independently of the chain length. I would write the following.} 
 If the infinitely long half chains were fully separated from each other already before the cut, these probabilities would be just 75\% and 25\% demonstrating that the states of the topological end spins of the left half chain are independent of each other leading to a triplet-singlet degeneracy in the ground state of the half chain. However, if the exchange coupling between the two half chains is finite before the cut, there is a small probability that virtual quasiparticles generated by vacuum-fluctuations cross the cut and alter the total spin of the half chain, although the probability of these processes is less than about 2\%. 

Semi-semiclassics can be used to determine the second contribution, that of pairs of quasiparticles created by the 
$J_2$ quench. This  gives rise to the quasiparticles' spin   distribution, $P_{\mathrm{qp}}(S_{\mathrm{qp}})$.
Assuming that  the spin orientation of the quasiparticles is independent of those of the vacuum fluctuations, we obtain  the total half chain spin distribution,
\begin{equation}
\label{eq:P(S,t)}
 P(S,t) = \sum_{S_{\mathrm{0}}} \sum_{S_{\mathrm{qp}}} \delta_{S \in S_0 \otimes S_{\mathrm{qp}} } \left(2S + 1\right) \frac{P_{\mathrm{0}}(S_{\mathrm{0}})}{2 S_{\mathrm{0}} + 1}
  \frac{P_{\mathrm{qp}}(S_{\mathrm{qp}};t)}{2 S_{\mathrm{qp}} + 1} \; .
\end{equation}

The distribution $P_{\mathrm{qp}}(S_{\mathrm{qp}})$ can, in general, be determined only numerically. However, 
as described in the next two sections, we have simple analytical expressions at very short times as well as  
 in the limits of completely reflective and  completely transmissive scatterings, valid for very cold and  very hot gases of quasiparticles, respectively. 

\begin{table}
\begin{center}
%\begin{tabular*}{\columnwidth}{|c|c|}
\begin{centering}
\begin{tabular}{|c|c|}
\hline
 \phantom{aa} $S_0$ \phantom{aa} & \phantom{a} $P_{0}(S_0)$ \phantom{a}\\
 \hline \hline
 0 & 0.2426 \\
 \hline
 1 & 0.7388 \\
 \hline
 2 & 0.0186 \\
 \hline
\end{tabular}
\end{centering}
%%%%%
\hskip 0.5cm
%%%%%
\begin{tabular}{|c|c|c|}
 \hline
 \phantom{aa} $S$ \phantom{aa} &  $(r_S/r_0)_{\rm SC}$ &  $(r_S/r_0)_{\rm TEBD}$ \\
 \hline
 0 & 1 & 1 \\
 \hline
 1 & 1.535 & 1.549(22)\\
 \hline
 2 & -2.481 & -2.502(20)\\
 \hline
 3 & -0.054 & -0.0572(9)\\
 \hline
\end{tabular}
\end{center}
\caption{Left: Ground state spin structure of a semi-infinite spin-1 Haldane chain. Right:   Slopes of the probability distributions at short times, $r_S = {\rm d}P(S,t)/{\rm d}t|_{t=0}$, as predicted by the semiclassical theory, and as extracted from the TEBD simulations for quench duration $t_Q=1.6/J_1$.}
\label{tab:P0}
\end{table}

\section{Short time ballistic behavior}\label{sec:Short_times}

First we analyze the initial change of $P(S)$ shortly after the quench, where the quasiparticle picture gives simple predictions. 
As discussed before, the initial state is a superposition of states containing randomly localized spin singlet quasiparticle pairs with random velocities $\pm v$, and a quasiparticles density $\rho$. The distribution of the magnitude of the velocity, $f(v),$ depends on the details of the quench.  The total spin of the left half-chain changes when the first $S=1$ quasiparticle crosses the position of the cut and carries spin from one half to the other. The probability of this happening within time $t$ is simply 
\begin{equation}
\label{eq:Qtau}
Q(t) = \rho \, \int_0^c vt f(v) dv \equiv \frac{t}{2 \tau}  \;,
\end{equation}
because the quasiparticle of velocity $v$ can come from an interval of length $|v|t$ on either side, touching the cut, but it must move to the right direction. The collision time $\tau$ above is defined as the ratio of the mean inter-particle spacing and the mean velocity.

For short times, we can neglect multiple crossings, so in Eq. \eqref{eq:P(S,t)} 
\bes
\label{eq:Pqp_short}
\begin{align}
P_\text{qp}(S_\text{qp}=1;t)&=t/(2\tau)+\dots \;,\\
P_\text{qp}(S_\text{qp}=0;t)&=1-t/(2\tau)+\dots\;,
\end{align}
\esu
and we find
%
%\bes
%\begin{align}
\begin{eqnarray}
&&P(0,t) = P_0(0)-\left(P_0(0)-\frac{P_0(1)}9\right)\frac{t}{2 \tau} +\dots \;,
\nonumber
\\
&&P(S,t) = P_0(S)
\label{eq:PS_short}
\\
&&\phantom{\;}+\frac{2S+1}6 \left(\frac{P_0(S-1)}{2S-1}-\frac{2\,P_0(S)}{2S+1}+\frac{P_0(S+1)}{2S+3}\right)\frac{t}\tau + \dots\,
\nonumber
%\end{split}
\end{eqnarray}
%\end{align}
%\esu
%

Using the vacuum spin probabilities in Table \ref{tab:P0}, we can compute, independently of $f(v)$, the ratios of the slopes $r_S = \left. \frac{d P(S,t)}{dt} \right|_{t=0}$ of the initial linear time dependences with the result shown in  Table~\ref{tab:P0}.

\begin{figure}
 \includegraphics[width = 0.5 \textwidth]{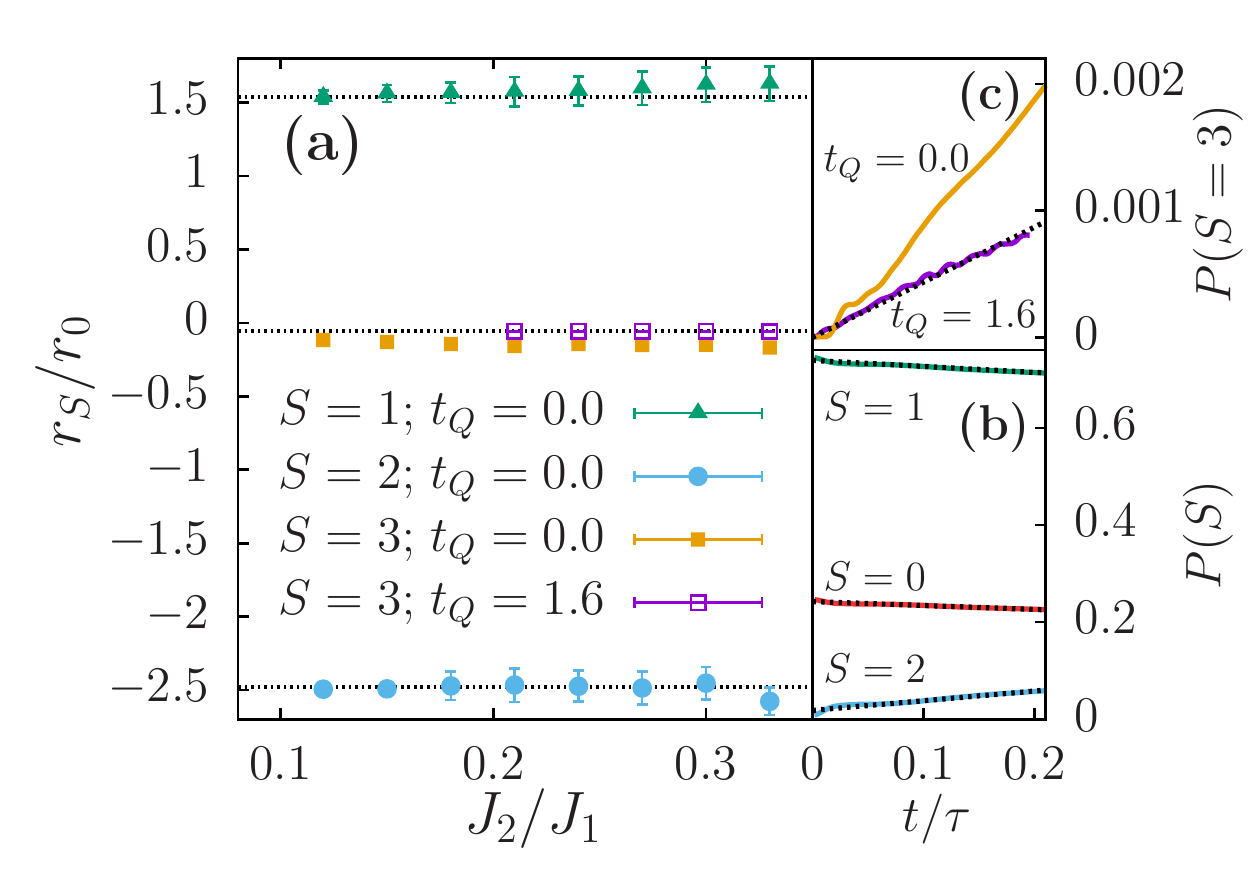}
 \caption{\textit{(Color online)} \textbf{(a)} The initial relative rates $r_S/r_0$ extracted from the microscopic simulation of the Heisenberg chain as a function of the quench magnitude, $J_2/J_1$. 
 Data for sudden quench are shown as symbols, while the dotted lines indicate the semi-semiclassical predictions. 
 The relative rates for a smooth quench ($t_Q = 1.6/J_1$)  are also plotted for $S=3$  as empty squares. 
%  The values for $S=1$ and $S=2$ agree well with 
% the theoretically predicted ones. In the case of $S=3$ the sudden quench data differs significantly from the predicted value, while for a smooth quench we find good agreement. 
% Due to strong transient quantum oscillations in the $S=3$ sector, it is hard to extract the universal short-time ($t \ll \tau$) rate in the case of a sudden quench (See inset (c)).
 \textbf{(b)} Short-time behavior of the $P(S,t)$ spin distribution for sectors $S=\lbrace 0,1,2 \rbrace$ in the case of a sudden quench $J_2/J_1=0.12.$ 
 Dotted black lines show linear fits used to determine the initial rates. \textbf{(c)} Short time behavior of  $P(S=3,t)$  for a sudden quench with $J_2/J_1 = 0.12$ and a smooth quench with $t_Q = 1.6/J_1$ and $J_2/J_1=0.33.$
 }\label{fig:rates}
\end{figure}

We have confronted  these predictions  with microscopic TEBD simulations, shown in Fig. \ref{fig:rates}. 
The short time $P(S,t)$ functions are plotted for $S=0,1,2$ in panel (b)  as functions of time, for a representative sudden quench. 
The relative rates $r_S/r_0$ extracted from these and similar curves obtained with various quench sizes are displayed in the main panel (a).  In accordance with the quasiparticle picture, the relative rates are independent of the quench size. The agreement with the quasiparticle prediction is excellent for $S=0,1,2,$ while there is a small deviation for sudden quenches for $S=3.$ The latter can be attributed to the difficulty of extracting the universal initial rate due to transient oscillations (c.f. upper curve in panel (c), comparing the results for  a sudden and a finite time quench for $S=3$). The oscillations are not present for smooth finite time quenches, and for these we get excellent agreement with the semiclassical prediction. This agreement between the prediction of the quasiparticle picture and the numerics gives strong evidence that the quasiparticle picture is correct. 
In the following sections we test the validity of the semiclassical description at longer times.

%\begin{equation}
% P_{\mathrm{cross}}(N;t) = \frac{Q(t)^N}{N!} e^{-Q(t)} \; ,
%\end{equation}
%where $Q(t)$ is the average number of the world lines that crossed the cut until time $t$ from both left and right,
%
%
%\begin{eqnarray}
%  P_{\mathrm{qp}}(S_{\mathrm{qp}}=0,t) &\approx& 1 - \frac{t}{2 \tau} + \mathcal{O}((t/\tau)^2) \nonumber \\
%   P_{\mathrm{qp}}(S_{\mathrm{qp}}=1,t) &\approx& \frac{t}{2 \tau} + \mathcal{O}((t/\tau)^2)
%\end{eqnarray}

%\begin{eqnarray}
% r_0 &=& - \frac{1}{2 \tau} P_{0}(0) + \frac{1}{18 \tau} P_{0}(1) \nonumber \\
% r_{S > 0} &=& - \frac{1}{3 \tau} P_{0}(S) + \nonumber \\
%        & & \frac{2S+1}{6 \tau} \left(\frac{P_0(S-1)}{2S-1} + \frac{P_0(S+1)}{2S+3} \right)
%\end{eqnarray}

\section{Collision dominated regime}\label{sec:Long_times}

At later times after the quench, several particles can cross from one half-chain to the other from both directions. Moreover, one has to take into account the effect of collisions. In this section, after considering two analytically tractable limiting cases, we apply the semi-semiclassical method and compare its results with the TEBD numerics.

\subsection{Simple limits}
\label{sub:simple}

Let us first consider two limits, those of completely reflective and completely transmissive collisions, in which we can  compute  the spin distribution function analytically.

\subsubsection{Completely reflective limit}

\begin{figure}[b]
 \includegraphics[width = 0.35 \textwidth]{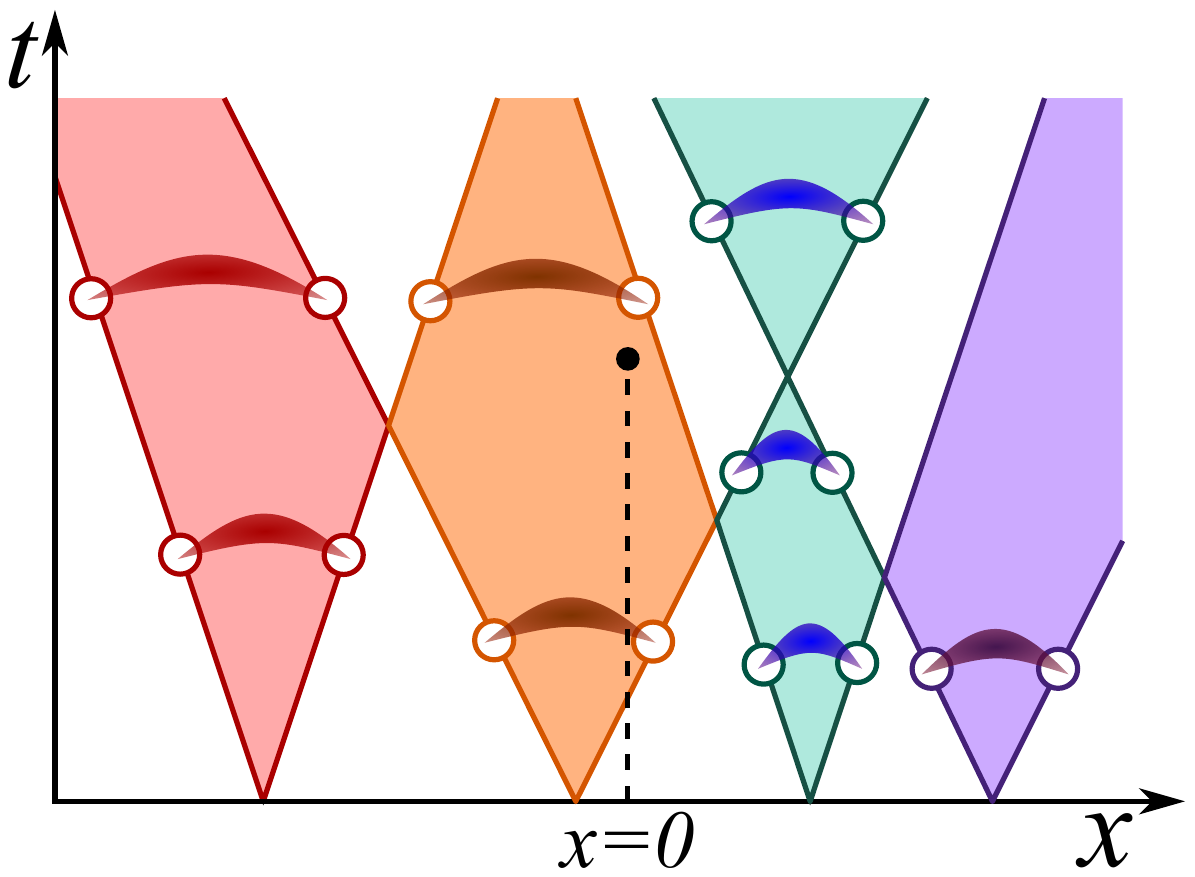}
 \caption{\textit{(Color online)} Visualization of quasiparticle spin dynamics in the fully reflective limit. 
 Initially the pairs form singlet pairs, indicated by colored shading. In this limit, singlet bonds between neighboring particles, indicated by arcs, remain intact.}
 \label{fig:reflective}
\end{figure}

In the universal low-energy limit,  the two-particle scattering matrix of gapped models with short range interaction is a permutation matrix, corresponding to perfect reflection of the incoming quasiparticles. This limiting S-matrix was used in the early works~\cite{SachdevDamle1997,Damle2005, RappZarand,Evangelisti2013,Kormos_Zarand_PRE_2016} on the semiclassical method to describe the dynamics at low temperatures. The S-matrix for the $O(3)$ nonlinear sigma model, describing our spin Hamiltonian, is  exactly known,  and also describes perfectly reflective processes  in the limit of small relative rapidities (see Appendix \ref{appendix:Smatrix}).

The initial state consists of pairs of quasiparticles that form spin singlets, and in this reflective limit,  neighboring quasiparticle pairs remain singlets even after many collisions (see Fig. 5). If we cut the system into two half-chains, the quasiparticle contribution to the total spin of the left part at a given time is $S_\text{qp}=0$ if the cut lies between pairs, while $S_\text{qp}=1$ if the cut breaks a pair. It is easy to see that the first situation is realized if the number of quasiparticles crossing from one half to the other up to the given time is even, and the second if this number is odd. The crossing number can be computed using the straight lines in Fig. 5 (the would-be trajectories of non-interacting particles). The total number of crosses from the left and right, $n_{+}$ and $n_{-}$,  follow independent  Poisson distribution, 
\begin{equation}
 p(n_{+},n_{-}) = \frac{1}{n_{+}! \, n_{-}!} q(t)^{n_{+} + n_{-}} e^{-2q(t)} \; ,
\end{equation}
where $q(t) = \frac12\rho \,t \int_0^c dv v f(v) = Q(t)/2=1/(4\tau)$ is the probability that a trajectory crosses the cut from the left (or from the right). 
%is the expected number of crosses from one side. 
The probabilities of having even ($S_\text{qp}=0$) or odd ($S_\text{qp}=1$) number of crossings are then
\begin{multline}
 P^\text{refl}_\text{qp}(S_\text{qp},t) = %p(n_{+}+n_{-} = \mathrm{even}) =
 \sum_{n_+,n_-=0}^\infty p(n_+,n_-)\frac{1+(-1)^{S_\text{qp}}(-1)^{n_++n_-}}2 \\
% =  \frac{1}{2}(1+(-1)^{S_\text{qp}}e^{-4q(t)}) 
= \frac{1}{2}(1+(-1)^{S_\text{qp}}e^{-t/\tau}) \;.
\end{multline}
%where $\tau^{-1} = 2 \rho \int_0^c dv \, v f(v)$ is the collision time. 
Note that expanding for short time we recover the expressions in Eqs. \erf{eq:Pqp_short}. Substituting the above result into Eq. \erf{eq:P(S,t)} the total spin distribution can be computed in the totally reflective limit.

\begin{figure} 
 \includegraphics[width = 0.35 \textwidth]{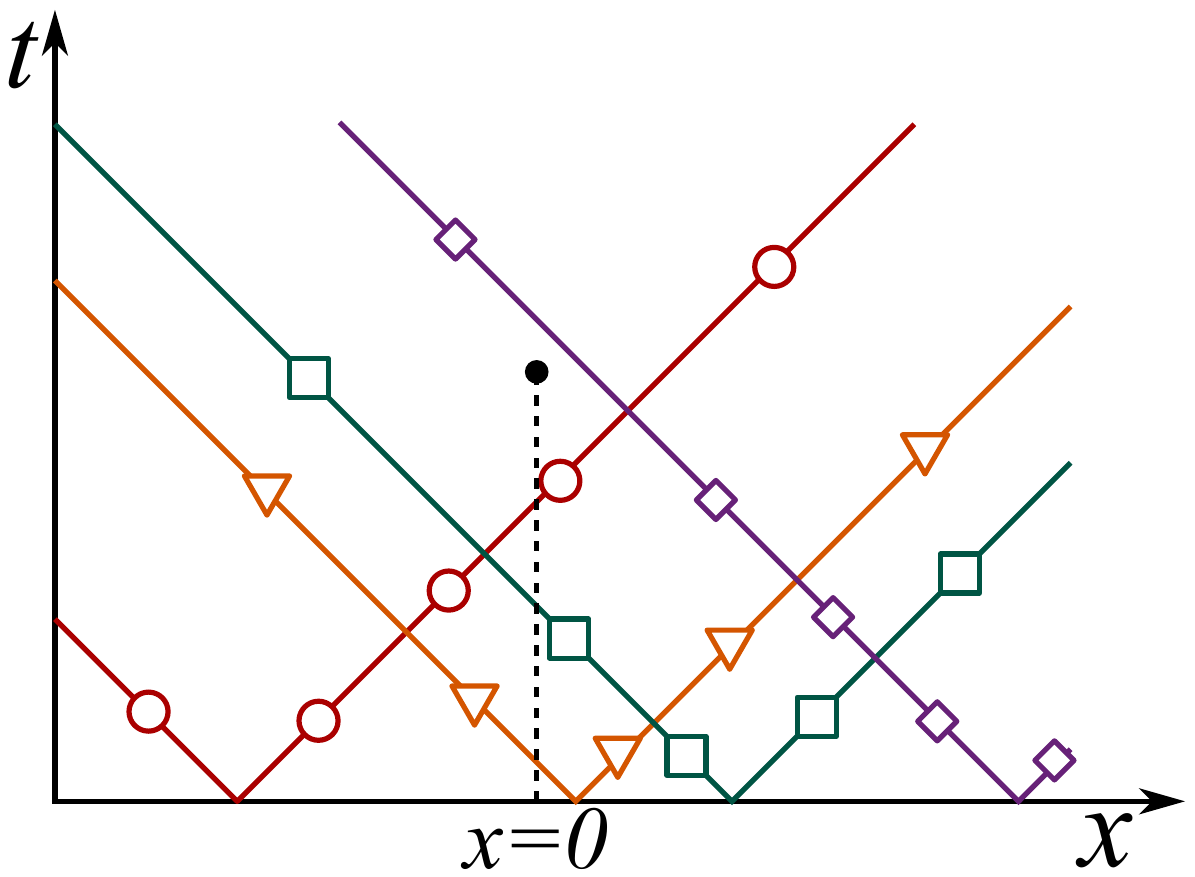}
 \caption{\textit{(Color online)} Quasiparticle spins in the fully transmissive (ultrarelativistic) limit.  Members of entangled quasiparticle pairs are marked by specific symbols on the worldlines. 
 Interaction between quasiparticles is negligible in this limit: the particles simply cross each other without changing their spins. }\label{fig:ultrarel}
\end{figure}

%\begin{eqnarray}
% P_{\mathrm{qp}}^{\mathrm{refl}}(S_{\mathrm{qp}}=0; t) = \frac{1}{2} + \frac{1}{2} e^{-t/\tau} \nonumber \\ 
%  P_{\mathrm{qp}}^{\mathrm{refl}}(S_{\mathrm{qp}}=1; t) = \frac{1}{2} - \frac{1}{2} e^{-t/\tau}
%\end{eqnarray}

\subsubsection{Completely transmissive (ultrarelativistic) limit}

Very high energy quasiparticles do not interact with each other.  This can also be verified on the exact S-matrix of the nonlinear sigma model in Appendix \ref{appendix:Smatrix},  in  the limit of very large rapidity differences, i.e. of ultrarelativistic quasiparticles.
In this limit, the original spin singlets remain singlets, but now the members of a pair keep moving away from each other following the light cone (see Fig. \ref{fig:ultrarel}). Since quasiparticles crossing the cut are independent of each other, the quasiparticle spin distribution is given by
\be
 P_{\mathrm{qp}}^{\mathrm{trans}}(S_{\mathrm{qp}}; t) = \sum_N
 P_{\mathrm{spin}}(S_{\mathrm{qp}}| N)   \;P_{\mathrm{cross}}(N;t) \; ,
\ee
where $P_{\mathrm{cross}}(N;t) = \frac{Q(t)^N}{N!} e^{-Q(t)}$ is the Poisson probability distribution of the number of worldlines that cross the cut from any side, while
$P_{\mathrm{spin}}(S|N) = (2S + 1)\, M_{S|N}/3^N$ is the distribution of the total spin of $N$ particles with random spin orientations. Here $M_{S|N}$ counts the 
multiplets of spin $S$ in the $N$-particle space, and it  can be calculated iteratively from the recursion relation
\be
M_{S|N} = M_{S+1|N-1} +(1-\delta_{S,0})\left(M_{S|N-1} + M_{S-1|N-1} \right)
\ee
with initial condition $M_{S|0} = \delta_{S,0}$. Solving these  equations iteratively, we can quickly compute 
$ P_{\mathrm{qp}}^{\mathrm{trans}}$ at any time.

\begin{figure*}[t]
 \includegraphics[width = 0.9 \textwidth]{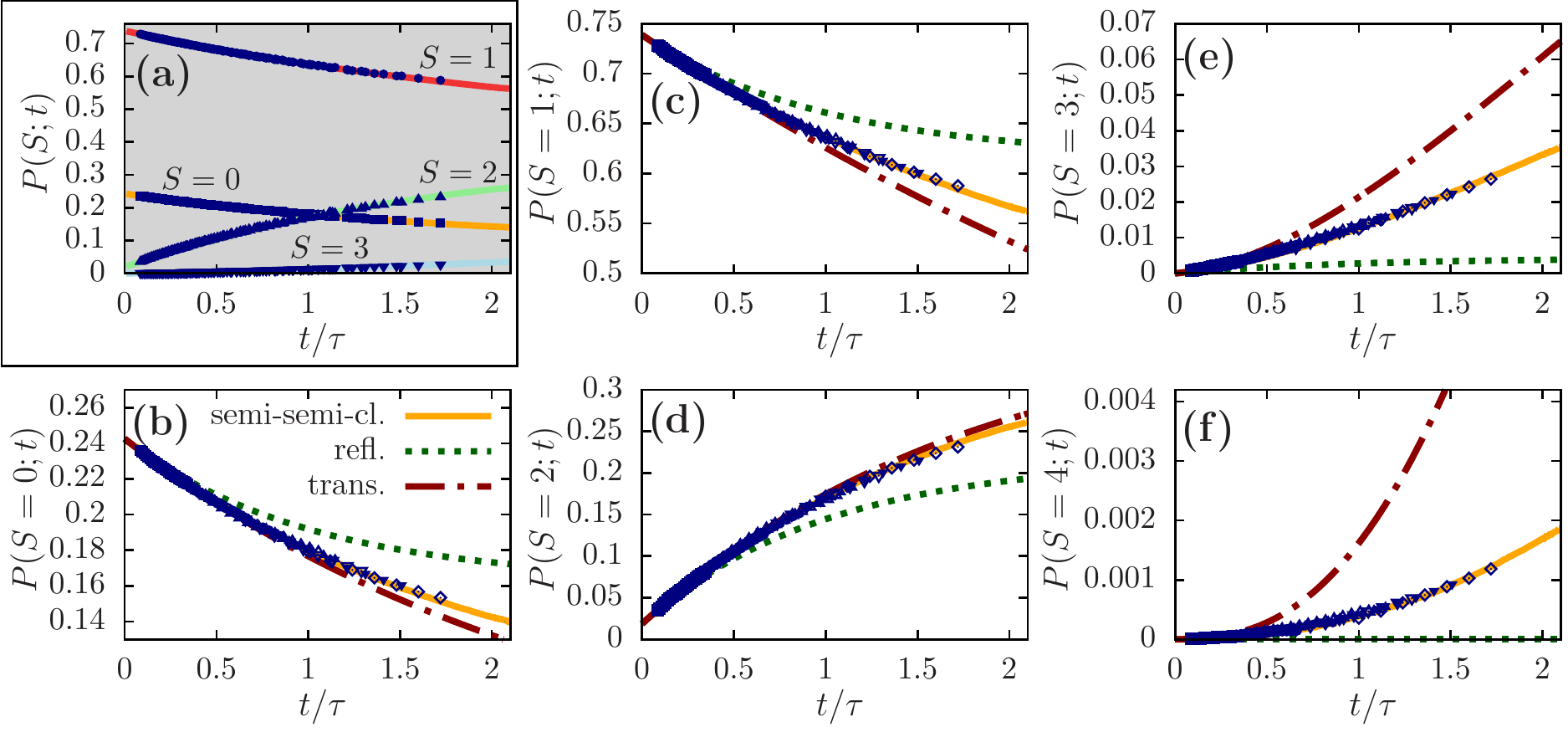}
 \caption{\textit{(Color online)} Spin distribution after a sudden quench as a function of $t/\tau$. Symbols are results of  microscopic simulations on the Heisenberg chain for different values of the 
 quench magnitude $J_2/J_1 \in [0.12, 0.36].$ The values of $\tau$ for different quench magnitudes were found by searching the best collapse of the curves. Panels (b-f) display the probabilities of each spin separately. The green dotted curves show the 
 prediction in the fully reflective limit, while  red dash-dotted lines show the fully transmissive values. The results of the microscopic simulation differ significantly from both limits, 
 but are well described by the semi-semi-classical simulation performed using a cut-off parameter $E_0 = 4J_1.$ Panel (a) shows 
 the curves $P(S;t)$  for all spin sectors together. Here  colored lines show the corresponding semi-semi-classical simulation. }\label{fig:fullTEBD}
\end{figure*}

\subsection{Hybrid semiclassical  dynamics}

%In this section we compare the results of the microscopic TEBD simulation for $P(S,t)$ with the semi-semiclassical method.

In general, the scattering is neither fully reflective nor fully transmissive. The result of each collision is instead a superposition of possible outgoing states with respective amplitudes given by the scattering matrix. In the O(3) nonlinear sigma model the total spin and also its $z$-component are conserved in the scattering of the $S=1$ quasiparticles, but transmissive and reflective processes as well as quasiparticle spin 
flips all  occur with finite scattering amplitude. In  the $S_T^z=0$ scattering channel, for example, a superposition of transmissions and reflections occur for incoming particles $(+,-)$, but even the process $(+,-)\longrightarrow(0,0)$ is allowed.

The two-body scattering  matrix  is exactly known for the O(3) nonlinear sigma model.
 In our hybrid semiclassical method we use this S-matrix: whenever there is a collision of quasiparticles, we act on the two colliding quasiparticle spins by the corresponding O(3) S-matrix. This goes beyond  standard semiclassical treatments not only by allowing nontrivial scattering processes, but also by treating the spin part of the many-body wave function fully 
 quantum mechanically \cite{PascuMarciGergo2017}.

Apart from the S-matrix, the other main input for the method is the momentum distribution of the quasiparticles, which is not easy to measure or calculate. In our simulation we used the distribution 
\be
\label{eq:p(q)}
n(q)\propto \frac{q^2}{\sinh^2(\omega_q/2 E_0)}\,,
\ee
with $\omega_q=\sqrt{\Delta^2+q^2c^2}$ the quasiparticle dispersion relation and $E_0$  a tunable cutoff parameter. This particular functional form is motivated by the perturbative calculation outlined in Sec.~\ref{sec:pert}. 

We compare the results of the semi-semiclassical method with the microscopic TEBD numerics for sudden quenches in Fig. \ref{fig:fullTEBD}.  Different symbols correspond to quenches of different magnitude.  In the semi-semiclassical approach, the distribution 
of scattering matrices depends exclusively on the velocity distribution of quasiparticles, which is, in turn, determined exclusively by the microscopic quench protocol. In the perturbative limit, we expect  this distribution to be independent of the amplitude of the quench. The amplitude of the quench is only supposed to influences the density of the  quasiparticles created, and thus the collision time. 
Therefore, we expect that  the functions $P(S,t)$ can be scaled on the top of each other by  rescaling  time.

This is indeed what we find by studying  sudden quenches of different sizes, $J_2/J_1 \in [0.12, 0.36]$, 
for which the curves $P(S,t)$  collapse  when plotted against $t/\tau$ (see Fig. \ref{fig:fullTEBD}).
 For smaller quenches, the collision time $\tau$  is determined from the early time slope of the functions in Eq. \erf{eq:PS_short}, while for larger quenches, where  transient oscillations are superposed at early times, we just rescaled the curves to achieve the best collapse. Remarkably, the same rescaling factor worked for all different spin values in each case.

The resulting curves can be compared to  semiclassical predictions. The first observation is that neither of the special limits can reproduce the results of the microscopic numerical simulation (c.f. green dotted and red dash-dotted curves in panels (b)-(f)). In particular, the standard semiclassical method based on totally reflective collisions cannot account for the post-quench dynamics,
and the numerically determined time evolution lies between the predictions of  completely reflective and completely transmissive computations. 

Remarkably, however,  the semi-semiclassical results give an excellent agreement with the TEBD data. We recall that there is a single free parameter in the semiclassical calculation, the cutoff parameter $E_0$. 
The same choice, $E_0=4J_1,$ gave the best agreement for all spin values $S$.
 This cutoff is close to the quasiparticles' bandwidth\cite{AffleckWhite2008}. 

\begin{figure}[b]
 \includegraphics[width = 0.48 \textwidth]{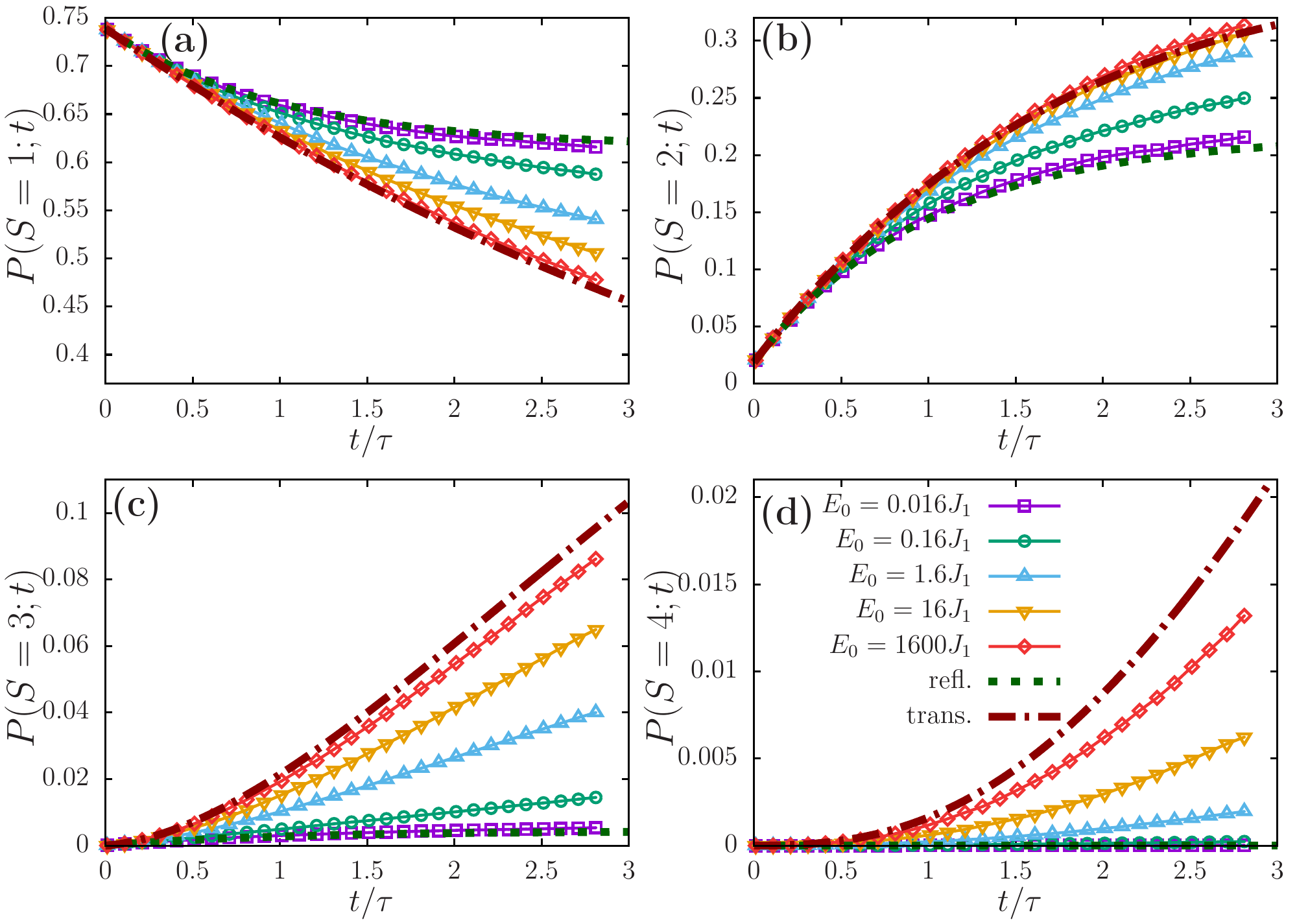}
 \caption{\textit{(Color online)} Spin distribution in the semi-semi-classical simulation with different cut-off parameters. The dashed green lines show the fully reflective limit, while the 
 red dash-dotted line shows the transmissive limit. The semi-semi-classical simulation results in spin distributions between these two limits, depending on the value of the cut-off. For small cut-off 
 the results are close to the reflective limit, while for large cut-off the transmissive limit is approached.
 }\label{fig:semisemi}
\end{figure}

The effect of the cutoff on the semi-semiclassical spin distributions for $S=1,\dots,4$ is illustrated in Fig. \ref{fig:semisemi}. A very small cutoff leads to slow quasiparticles with predominantly reflective scattering, and dynamics  close to the standard semiclassical prediction of Subsection~\ref{sub:simple}. For large cutoffs, on the other hand,  the majority of the quasiparticles is ultrarelativistic and collide mostly by  transmissive scatterings. Varying $E_0,$ the semi-semiclassical approach interpolates between these two limiting cases.

\section{Perturbative quench theory}
\label{sec:pert}
In this section, we develop a perturbative description of the quench in terms of an effective field theory, 
and compare its implications with the results of our TEBD simulations and the semi-semiclassical interpretation thereof. 
Certain details of the derivations are given in Appendix \ref{appendix:pert}.

Our first step is to replace the original Hamiltonian \erf{eq:Hamiltonian} by a phenomenological quasiparticle 
Hamiltonian, and to express the post-quench Hamiltonian as %
\begin{equation}
 \hat{H}_0 = \sum_{q,\sigma} \omega_{q}  \; b^{\dag}_{q,\sigma} b_{q,\sigma} \; .
 \label{eq:H_eff}
\end{equation}
Here the operators  $b^\dagger_{q,\sigma}$ create  quasiparticles of momentum $q$, energy 
$\omega_q$, and spin $\sigma =\pm1,0 $.  As discussed earlier, switching off $J_2$ creates singlet pairs of quasiparticles with opposite momenta. This is implemented 
within our effective field theory by the 
term\cite{footnote_effective_Hamiltonian}
\begin{multline}
\label{eq:Hquench}
 \hat{H} = \sum_{q,\sigma}  \left\lbrace \omega_{q}(t) b^{\dag}_{q,\sigma} b_{q,\sigma} + \phantom{\frac{a}{b}}\right. \\
  \left.\frac{1}{2} g_q(t) (-1)^\sigma \left(b^{\dag}_{q,\sigma} b^{\dag}_{-q,-\sigma} +
 b_{q,\sigma} b_{-q,-\sigma} \right) \right\rbrace\,,
\end{multline}
with the time dependent dispersion and pair creation amplitudes 
\begin{equation}
  \omega_{q}(t) = \omega_q + J_2 \lambda(t) \, \Delta_q \; , \quad g_q(t) = J_2 \lambda(t) \, g_q\,.
\end{equation}
Here $\lambda(t)$ describes the time-dependent profile of the quench with $\lambda(t \rightarrow -\infty) = 1$ and $\lambda(t \rightarrow \infty) = 0.$ The precise momentum dependence of the couplings  $\Delta_q$ and $g_q$ depends on  microscopic details of the Hamiltonian $\hat{H}_0$ and the perturbing quench operator. At leading order in perturbation theory, the momentum distribution of quasiparticles is determined by the quasiparticle dispersion $\omega_q$, the specific form of $\lambda(t)$, and by the pair creation amplitude, 
 $g_q$. This latter typically vanishes in the limit  $q\to 0$ linearly, reflecting the fact that low energy quasiparticles behave  as 
 hard-core bosons.

Being quadratic,  the simple model outlined above can be treated analytically, and can be used to determine 
 the momentum distribution of the quasiparticles analytically. For small and smooth quenches we find
\be
n(q)\propto |g_q|^2 \; \frac{J_2^2}{\omega_q^2}\;\; \biggl| \widetilde{\frac{{\rm d} \lambda}{{\rm d}t}}  (2\omega_q)\biggr|^2\,,
\label{eq:n(q)}
\ee
where tilde denotes the Fourier transform. Note that $\lambda(t)$ does not depend on the amplitude of the quench, only on its duration and shape. Quite naturally, the quasiparticle density is proportional to the squared magnitude of the quench. Consequently, the  collision rate $1/\tau$ defined in Eq. \erf{eq:Qtau} is also just proportional to $J_2^2$. 

\begin{figure}[t]
 \includegraphics[width = 0.4 \textwidth]{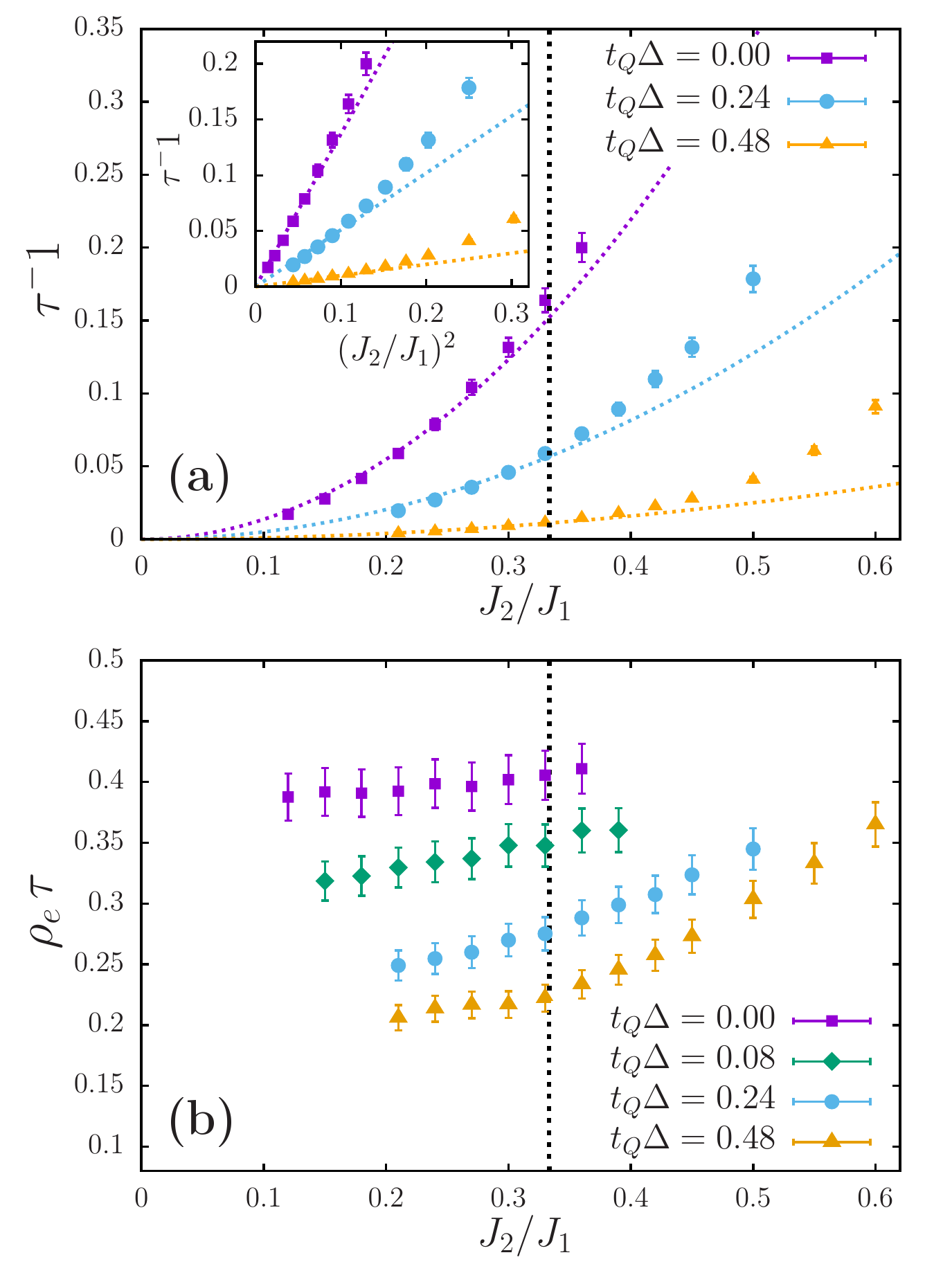}
 \caption{\textit{(Color online)} Checking basic predictions of perturbation theory. Different symbols show data for different quench times $t_Q.$ \textbf{(a)} The fitted inverse collision time $\tau^{-1}$ as a function of the quench magnitude $J_2/J_1$. 
 The vertical dashed line indicates the AKLT point, $J_2/J_1 = 1/3.$ Perturbation theory predicts quadratic dependence of $\tau^{-1}$ on $J_2/J_1$ that seems to be valid up to the AKLT point.
 The inset shows the same data but with quadratic scale on the horizontal axis. \textbf{(b)} The product of the energy density $\rho_\text{e}$ and the collision time $\tau$ as a function of
 the quench magnitude $J_2/J_1.$  Perturbation theory predicts this value to be constant for a given value of $T_Q,$ which seems to hold up to the AKLT point. The error bars arise from the uncertainity of the fitted value of $\tau$, while the relative error of the measured energy density is small.}\label{fig:perturbation}
\end{figure}

This prediction is checked in Fig. \ref{fig:perturbation}.a, where we plot the  inverse collision times extracted from TEBD as  functions of $J_2/J_1$ for three different quench speeds. The quadratic dependence is valid up to $J_2/J_1\approx1/3$ which is just the so-called AKLT point \cite{AKLT}, suggesting that perturbation theory holds for quenches even up to this size. Also, slower, more adiabatic  quenches create less quasiparticles, reflected in a lower collision rate.

As the quench magnitude $J_2$ only appears in the prefactor in Eq.~\eqref{eq:n(q)}, 
the normalized velocity distribution is predicted to be independent of it in this perturbative regime. 
 The energy density should therefore be proportional to the quasiparticle density with a proportionality factor depending on the shape of $\lambda(t)$.  As a consequence, the product of the collision time and the energy density $\rho_\text{e}$ should  be independent of the size of the quench if perturbation theory holds. This prediction is tested in Fig. \ref{fig:perturbation}(b), where we plot the product $\rho_\text{e}\tau$ against $J_2/J_1$ for various quench durations. 
Here the quasiparticle energy density was measured in the TEBD simulation by taking the difference of the post-quench and the vacuum energy densities. It can be seen that the product is indeed approximately constant up to $J_2/J_1\approx1/3,$ i.e. in the domain of perturbation theory identified above.

%From \eqref{perturbative_momdist} the perturbative predictions:
%\begin{itemize}
% \item the momentum (velocity) distribution is independent of $J_2$, depends only on the shape of $\gamma(t)$.
% \item the quasiparticle density: $\rho \propto |J_2|^2 \Rightarrow \tau^{-1} \propto |J_2|^2$
% \item the energy density is $\rho_e = \eta_{\gamma} \rho$ where the coefficient $\eta_\gamma$ depends  only on the shape of $\gamma(t)$.
% \item For the energy density $\rho_{e} \tau = \mathrm{const.}$.
% \end{itemize}

\begin{figure}[b]
 \includegraphics[width = 0.4 \textwidth]{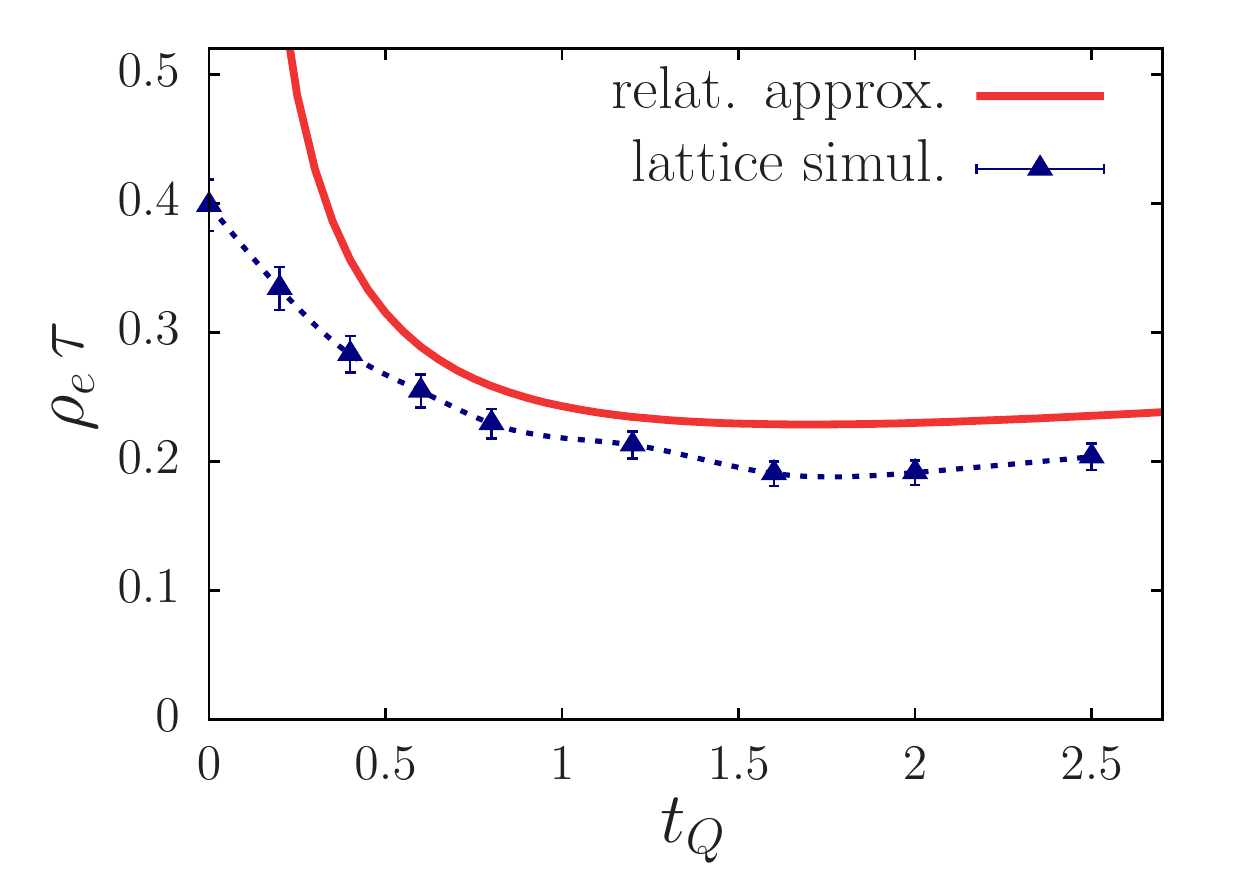}
 \caption{\textit{(Color online)} The product of the energy density $\rho_e$ and the collision time $\tau$ as a function of the quench time $t_Q$ for $J_2/J_1 = 0.24$. The continuous red line shows the 
 prediction of the relativistic minimal model. While the simple prediction largely fails for small quench times $t_Q$, for larger $t_Q$ the agreement is reasonable.}\label{fig:relat_test}
\end{figure}
 
In order to make a more quantitative comparison between  TEBD  and our semiclassical method, 
we  specify the shape of the quench as
\be
\lambda(t) = \frac{1}{2} \left( 1 - \tanh(t / t_Q) \right)\,.
\ee
We furthermore assume  a simple linear dependence of $g_q$ around $q=0$ (see Appendix \ref{appendix:pert}) and use a relativistic dispersion relation $\omega_q$  with the microscopic parameters given in Eq. \erf{eq:gapc}. With these approximations, we obtain 
 the quasiparticle momentum distribution used in our semi-semiclassical simulations, Eq.~\eqref{eq:p(q)} with $E_0 = t_Q$,
\be
\label{eq:1/sinh}
n(q) \propto J_2^2\frac{q^2}{\sinh^2(\pi t_Q \omega_q)}\,.
\ee
With this distribution, we can compute the product of the energy density and the collision time for different quench durations $t_Q.$ The results are shown in Fig. \ref{fig:relat_test} together with the TEBD data for  quenches with $J_2/J_1=0.24.$ There is a reasonable agreement for not very fast quenches. This demonstrates convincingly that the semiclassical picture based on relativistic quasiparticles with perturbatively computed momentum distribution provides a valid qualitative and quantitative description of the post-quench dynamics.

Let us close this Section by verifying the consistency condition \eqref{eq:lowdens} of the semi-semiclassical approach. 
Even though we cannot measure the particle density directly, we can set an upper bound on it   using the energy density:
\be\label{eq:endens_guess}
\rho\lesssim \rho_e/\Delta\,.
\ee
This implies that the semiclassical condition \eqref{eq:lowdens} is certainly satisfied if 
\be
\label{eq:cond}
c \rho_e < \Delta^2 
\ee
holds.  We checked that this inequality was always satisfied in our simulations, except for the largest sudden quench with $J_2/J_1 = 0.36$, where $c \rho_E / \Delta^2 \approx 1.2$ was found. However, for sudden quenches \eqref{eq:endens_guess} strongly overestimates the particle density, i.e. \eqref{eq:lowdens} is expected to be valid even in this largest sudden quench.

\section{Summary}\label{sec:summary}

In this work, we have studied numerically  the  statistics of spin transfer after a quantum quench in the Haldane chain, and compared it with the predictions  of the semi-semiclassical approach, developed in Ref.~\onlinecite{PascuMarciGergo2017}. 
As we demonstrate, the probability distribution $P(S,t)$ of transferred spins, extracted from non-Abelian TEBD quench  simulations, contains a surprising  amount of information: in addition to detecting the presence of topologically protected end states \cite{AKLT,Kennedy1990},
it gives insight to the internal spin structure and velocity distribution of quasiparticles as well as to the nature of their collisions. 

Application of SU(2) symmetries allows us to reach  sufficiently large bond dimensions in the range of 2000 multiplets corresponding to 10,000 states \cite{Werner_future}, and to reach long enough simulation times to overlap with the range of validity of semi-semiclassics, and to observe effects related to quasiparticle collisions. 
We find that  spin distributions, as computed through our full  TEBD simulations, are in perfect agreement with those determined from the semi-semiclassical approach: the short time behavior  of $P(S,t)$ reveals the presence of spin $S=1$ quasiparticles traveling ballistically, while at longer times we enter  a collision-dominated regime, where collisions are neither reflective~\cite{SachdevDamle1997,RappZarand,Evangelisti2013,Kormos_Zarand_PRE_2016} nor transmissive \cite{Altshuler2006}. In this regime, it appears to be absolutely necessary to  incorporate the coherent time evolution of the spin wave function, and to follow its quantum mechanical  evolution, as performed by the hybrid  semi-semiclassical  approach. 

For a sudden quench, the functions $P(S,t)$ are found to be universal functions of $t/\tau$, with  $\tau$  the collision time. This universal scaling can  also be simply explained within the semi-semiclassical theory: the size of the quantum quench has a direct influence on the quasiparticle density, $\varrho\sim J_2^2$, and thus the scattering rate, $1/\tau\sim \rho$, however, it does not change the velocity distribution of the quasiparticles and the structure and distribution of scattering matrices. Therefore, the statistics of collisions and the statistical time evolution of the wave function is independent of the size of the quench if time is measured in units of collision time.  

The precise velocity distribution does depend, however, on details of  the quench protocol and the quench time $t_Q.$ Correspondingly, for finite time quenches, we observe an explicit dependence 
of the  functions  $P(S,t)$  on the quench time (quench protocol). This explicit dependence is very well captured by simply changing the energy cut-off of the quasiparticles' velocity distribution, $E_0\sim 1/t_Q$.  This latter correspondence, and the velocity distributions used  in our semi-semiclassical simulations, have both been  motivated by a perturbative field theoretical quantum quench theory, presented in Section~\ref{sec:pert}. 

The extension of perturbative quench regime as well as the range of validity of our semi-semiclassical approach appear to be astonishingly large. First of all, $1/\tau$, directly proportional to the density of  quasiparticles is found to scale as $\sim J_2^2$ up to quench sizes of $J_2/J_1\approx 0.3,$ a value very close to the AKLT point at $J_2/J_1=1/3.$ Second, our assumption 
of a quench size independent velocity distribution is also  verified  by direct measurements of the quasiparticles' energy density on the Haldane chain.  Also quite remarkably,  our perturbative field theoretical calculation  assuming a simple relativistic quasiparticle spectrum yields a \emph{quantitative} estimate for the dimensionless energy density $\rho_e \tau\sim \rho_e/\rho$, in very good agreement with our full TEBD spin chain simulations for quench times $t_Q\gtrsim 0.5 /J_1 $. This is, again, quite astonishing, since sudden quenches generate many quasiparticles presumably outside the  regime of  our effective field theory. Apparently, however, quasiparticles close to the gap appear to play the dominant role in the spin transport studied here. 
 
The calculations presented here have thus two general conclusions: on the one hand, they show that semi-semiclassical calculations  are simple methods that are able to capture important long-time features of  non-equilibrium quantum dynamics at time scales much above the microscopic time scales.  On the other hand, our studies reveal the power and richness of full counting statistics and full distributions, such as $P(S,t)$, containing precious information on correlations, equilibration and dynamics \cite{Gustavsson2006, Lamacraft2008,Hofferberth2008,Batalhao2014,Izabella2017,Collura2017}.

\emph{Acknowledgements.} This  research  has  been  supported  by the Hungarian National Research, Development and Innovation Office (NKFIH) through Grant Nos. SNN118028, K120569, and the Hungarian Quantum Technology National Excellence Program (Project No.  2017-1.2.1-NKP-2017- 00001). C.P.M. aknowledges support from the Romanian National Authority for Scientific Research and Innovation (UEFISCDI) through Grant No. PN-III-P4-ID-PCE-2016-0032.  M.K. was also supported by a “Pr\'emium” postdoctoral grant of the Hungarian Academy of Sciences.  \"O. L. also acknowledges financial support from the Alexander von Humboldt foundation.

\clearpage
\newpage
%\pagebreak

\appendix

\section{NA-TEBD}\label{appendix:TEBD}
\begin{figure}
 \includegraphics[width = 0.5 \textwidth]{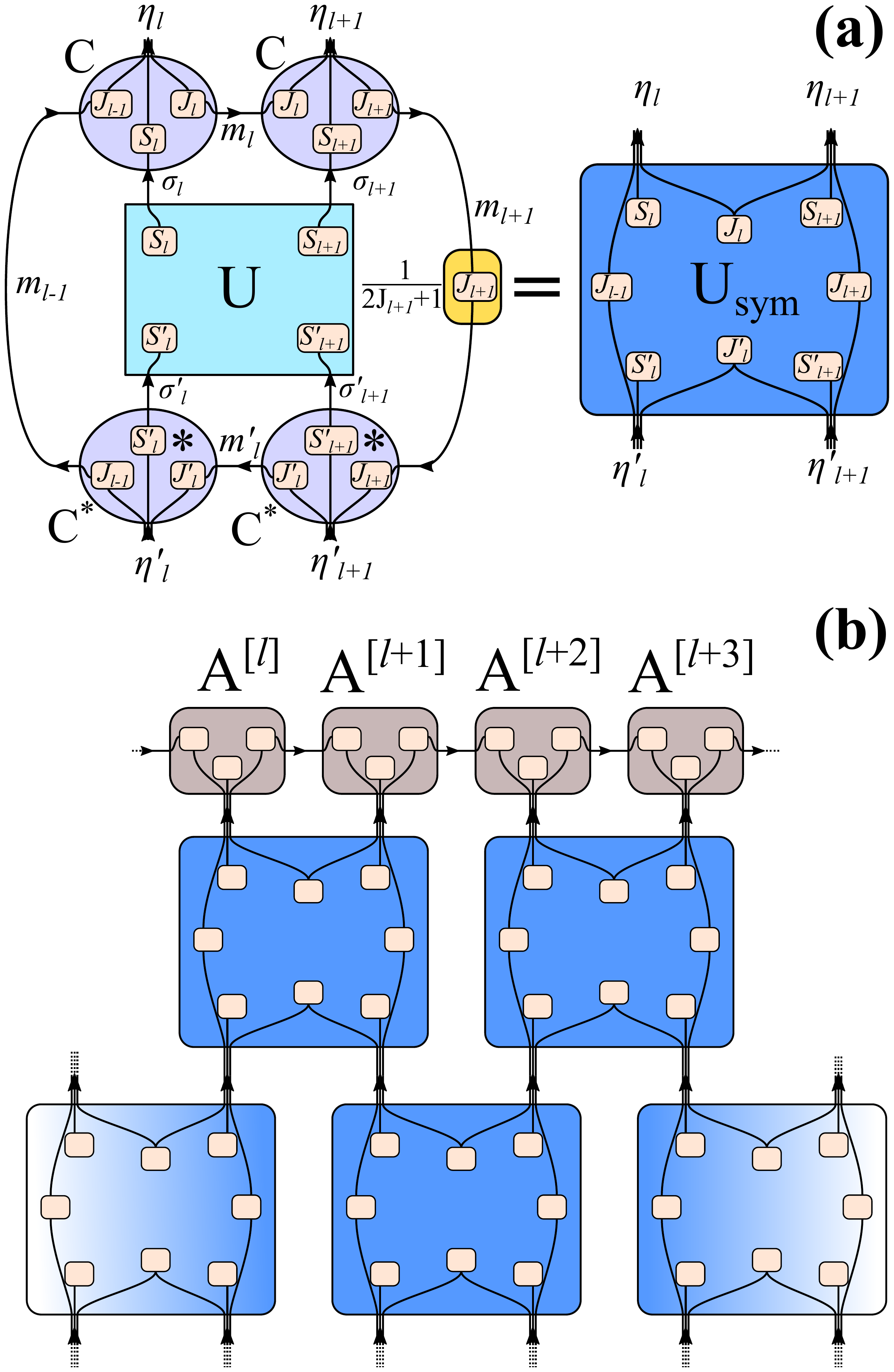}
 \caption{\textit{(Color online)} \textbf{(a)} Graphical representation of the defining equation \eqref{eq:NATEBD_evolver}. The blocks of the resulting $\left[U_{\mathrm{sym}} \right]_{\eta_{l}' \eta_{l+1}'}^{\eta_{l} \eta_{l+1}}$ tensor are specified by eight irreducible representation indices. \textbf{(b)} The NA-TEBD algorithm. The symmetric evolvers $U_{\mathrm{sym}}$ are directly applied on the upper layer of the NA-MPS. }\label{fig:NATEBD}
\end{figure}
The MPS based calculations of the time-evolution -- both the microscopic simulations of the $S=1$ chain and the hybrid semi-semiclassical simulations -- were performed by a variant of the standard TEBD algorithm~\cite{Vidal2007} that exploits the presence of the non-Abelian SU(2) symmetry; we call our variant of the algorithm NA-TEBD, and our code is similar to the approaches described in Refs.~\onlinecite{Singh_Vidal2010, Wilson_NRG, Pascu_SU3_NRG,Weichselbaum2012}, but by the introduction of the $\eta$-legs the generalization to other non-Abelian symmetries becomes easier~\cite{Werner_future, footnote_on_outer_multiplicity}.

In the TEBD algorithm, the time-evolution of the MPS wave function is simulated by consecutive application of two-site unitary evolvers~\cite{Vidal2007}, based on the Trotter--Suzuki decomposition~\cite{Trotter, Suzuki}. Considering now the SU(2) symmetric case, the evolvers like $U_{\sigma_{l}' \sigma_{l+1}'}^{\sigma_{l} \sigma_{l+1}}$, acting on the neighboring sites $l$ and $l+1$, preserve the SU(2) symmetry, and one naturally would like to exploit this property. Formally, one should contract the legs ${\sigma_{l}, \sigma_{l+1}}$ of the tensor $U$ with the ones of the Clebsch--Gordan tensors in the two-layer NA-MPS (See Fig.~\ref{fig:NAMPS} and Fig.~\ref{fig:NATEBD}). However, if the SU(2) symmetry is preserved by the two-site evolver, then after the application of it, the Clebsch-layer of the NA-MPS should remain the same, i.e. only the tensors $A^{[l]}$ and $A^{[l+1]}$ change. Using the standard orthogonality relations of the Clebsch--Gordan coefficients, we can determine the ``symmetric'' version of the two-site evolver that can be directly applied on the legs $\eta_l$ and $\eta_{l+1}$ of the tensors $A^{[l]}$ and $A^{[l+1]}$. This ``symmetric'' evolver is defined as (see also Fig.\ref{fig:NATEBD}a)
\begin{eqnarray}\label{eq:NATEBD_evolver}
 \left[U_{\mathrm{sym}} \right]_{\eta_{l}' \eta_{l+1}'}^{\eta_{l} \eta_{l+1}} =  \sum_{\substack{m_{l-1} \\ m_{l}\\ m_{l}' \\ m_{l+1}}} \sum_{\substack{\sigma_{l} \\ \sigma_{l+1} \\ \sigma_{l}' \\ \sigma_{l+1}'}}  \frac{1}{2 J_{l+1} + 1} U_{\sigma_{l}' \sigma_{l+1}'}^{\sigma_{l} \sigma_{l+1}} \nonumber \\ C^{m_l\, \eta_l}_{m_{l-1} \, \sigma_{l}} C^{m_{l+1}\, \eta_{l+1}}_{m_{l} \, \sigma_{l+1}}  \left(C^{m_l'\, \eta_l'}_{m_{l-1} \, \sigma_{l}'} \right)^{*} \left( C^{m_{l+1}\, \eta_{l+1}'}_{m_{l}' \, \sigma_{l+1}'} \right)^{*} \; .
\end{eqnarray}
The resulting tensor can be determined and stored before the simulation procedure, and later can be directly applied on the upper layer of the NA-MPS. It is important to remark that the blocks of the resulting tensor $\left[U_{\mathrm{sym}} \right]_{\eta_{l}' \eta_{l+1}'}^{\eta_{l} \eta_{l+1}}$ are specified by eight irreducible representation (``total-spin'') indices that are $\lbrace J_{l-1}, S_{l}, J_{l}, S_{l+1}, J_{l+1}, S_{l}', J_{l}', S_{l+1}' \rbrace$. 

After the determination of $U_{\mathrm{sym}}$ the standard TEBD algorithm as described in Ref.~\onlinecite{Vidal2007} can be used to evolve the upper layer of the NA-MPS without any relevant modification as shown in Fig.\ref{fig:NATEBD}b. In other words, the layer of Clebsch--Gordan coefficients does not appear during the simulations, making our NA-TEBD code fast and efficient.

\textit{Numerical details.} The rapidly growing truncation error (the weight of discarded Schmidt states) that spoils the accuracy of TEBD simulations leads to a short cutoff time beyond which results get unreliable. In our simulations we defined this cutoff time where the truncation error reached $10^{-7}$. To reach long enough times ($t \gtrsim \tau$), the MPS bond-dimension was set up to $M_{SU(2)} = 2000$, i.e. 2000 multiplets (around 10,000 states) with the largest Schmidt values were kept.

\section{The relativistic S-matrix}
\label{appendix:Smatrix}
In the hybrid semi-semiclassical simulations the relativistic S-matrix of the low-energy field theory of the $S=1$ Heisenberg model (namely, the $O(3)$ non-linear sigma model) was used to describe the collision events. We used the same S-matrix as in Ref.~\onlinecite{Marci_Pascu_Gergo_NESS}, but we transformed it in the more convenient form  
\begin{widetext}
\begin{equation}
 S = \left( \begin{array}{ccccccccc} 
\sigma_2 + \sigma_3 & 0 & 0 & 0 & 0 & 0 & 0 & 0 & 0 \\
0 & \sigma_3 & 0 & \sigma_2 & 0 & 0 & 0 & 0 & 0 \\
0 & 0 & \sigma_1 + \sigma_3 & 0 & -\sigma_1 & 0 & \sigma_1 + \sigma_2 & 0 & 0 \\
0 & \sigma_2 & 0 & \sigma_3 & 0 & 0 & 0 & 0 & 0 \\
0 & 0 & -\sigma_1 & 0 & \sigma_1 + \sigma_2 + \sigma_3 & 0 & -\sigma_1 & 0 & 0 \\
0 & 0 & 0 & 0 & 0 & \sigma_3 & 0 & \sigma_2 & 0 \\
0 & 0 & \sigma_1 + \sigma_2 & 0 & -\sigma_1 & 0 & \sigma_1 + \sigma_3 & 0 & 0 \\
0 & 0 & 0 & 0 & 0 & \sigma_2 & 0 & \sigma_3 & 0 \\
0 & 0 & 0 & 0 & 0 & 0 & 0 & 0 & \sigma_2 + \sigma_3 \end{array} \right) \; .
\end{equation}
\end{widetext}
Here the (incoming and outgoing) basis states are $\lbrace\ket{++}, \ket{+0}, \ket{+-}, \ket{0+}, \ket{00}, \ket{0-}, \ket{-+}, \ket{-0}, \ket{--} \rbrace$, where $\lbrace +,0,- \rbrace$ stand for the eigenvalues $S^z= \lbrace 1,0,-1 \rbrace$ of the particles. In the two-particle basis states we used the convention that the left and right spin-state always stand for the particles in the left and right position, both in the incoming and outgoing states. The values $\sigma_i$ depend on the relative rapidity $\theta_{\mathrm{rel}}=\theta_1-\theta_2$ of the colliding particles as
\begin{eqnarray}
 \sigma_1 &=& \frac{2 i \pi \theta_{\mathrm{rel}}}{(\theta_{\mathrm{rel}}+i \pi)(\theta_{\mathrm{rel}} - 2i \pi)} \; ,   \nonumber \\
 \sigma_2 &=& \frac{\theta_{\mathrm{rel}} (\theta_{\mathrm{rel}}-i \pi)}{(\theta_{\mathrm{rel}}+i \pi)(\theta_{\mathrm{rel}} - 2i \pi)}  \; ,   \nonumber \\
 \sigma_3 &=& \frac{-2 i \pi (\theta_{\mathrm{rel}}-i \pi)}{(\theta_{\mathrm{rel}}+i \pi)(\theta_{\mathrm{rel}} - 2i \pi)} \; .
\end{eqnarray}

\section{Perturbative description of the quench}
\label{appendix:pert}

In this appendix we provide some details on the quench dynamics within our effective field theory, presented in Section~\ref{sec:pert}.

We shall assume in what follows that our initial quench state is a squeezed state \cite{Fioretto2010,Horvath2016}. 
Using that our time dependent Hamiltonian \eqref{eq:Hquench} is quadratic, one can show that the wave function remains a squeezed state upon time evolution,
\begin{equation}
 \ket{\Psi(t)} \sim e^{\sum_{q>0,\sigma} \left(i \Phi_{q,\sigma}(t) + (-1)^\sigma K_{q}(t) b^\dag_{q,\sigma} b^\dag_{-q,\overline{\sigma}} \right)} \ket{0} \,.
\end{equation}
The exponential form corresponds to independent pairs of quasiparticles.
%and is known to be a good approximation of  states created through small quenches. 
%{\bf For fermionic quasiparticles $K_q$ must be an odd function of $q$ and it is natural to assume a linear behavior near the origin.}

Our task is to determine the full time evolution of the functions $K_{q}(t)$ throughout a quantum quench. 
First we observe that  different momentum modes decouple as  $\ket{\Psi(t)} = \prod_{q>0} \ket{\Psi_q(t)}$, with
\be
\ket{\Psi_q(t)} = \sqrt{1-|K_q|^2}e^{i\Phi_q} \sum_{n=0}^\infty \frac1{n!}K_q^n (b^\dag_q b^\dag_{-q})^n\ket{0}\,.
\ee
The time-dependent Schr\"odinger equation leads to an equation for the  pair creation amplitude $K_q,$
\begin{equation}
 i \dot{K}_q (t) = 2 \omega_q (t) K_q(t) + g_q(t) \left(1 + K_q(t)^2 \right) \; .
\label{eq:Kt}
\end{equation}
Within first order perturbation theory we can neglect the nonlinear $K_q(t)^2$ term in this equation. Using the initial values of the parameters, $\omega_q(t \rightarrow - \infty) = \omega_q + J_2 \Delta_q$ and $g_q(t \rightarrow - \infty) = J_2 g_q$, and exploiting that in the limit $t \rightarrow -\infty$ we initialize the system in its stationary vacuum state, we find the following initial conditions,
\be
\lim_{t\to-\infty}K_q(t) = -\frac{J_2g_q}{2(\omega_q+J_2\Delta_q)} = -\frac{J_2g_q}{2 \omega_q} +O(J_2^2) \,.
\ee

In general, one needs to solve Eq.~\eqref{eq:Kt} numerically. However, we can easily  obtain its approximate analytical solution 
up to leading order  in $J_2$, 
\be
\lim_{t\to\infty} K_q(t) = \frac{J_2g_q}{2\omega_q}e^{-2i\omega_qt}\int_{-\infty}^\infty dt' \dot\lambda(t')e^{2i\omega_qt'} +O(J_2^2)\,.
\ee
Using this expression we find that the density of quasiparticles created by the quench is just given by
Eq.~\eqref{eq:n(q)}. 

For the particular quench profile
\be
\lambda(t) = \frac{1}{2} \left( 1 - \tanh(t / t_Q) \right)\,,
\ee
the Fourier transformation can be performed analytically with the result
\be
n(q) = \frac{\pi^2}4 \frac{J_2^2g_q^2 t_Q^2}{4\sinh^2(\pi\omega_q t_Q)}\,.
\ee
%,
For slow enough quenches the denominator cuts off the momentum distribution at low momenta, so we can substitute $g_q$ with its small momentum behavior, $g_q\approx g\;|q|,$ and we obtain the momentum distribution \erf{eq:1/sinh}.

%\section{Semiclassical spin fluctuations}


\begin{thebibliography}{99}
\bibitem{footnote_debroglie} More precisely, one should consider the equivalent of the thermal de Broglie wavelength which is related to the momentum of the particles. We estimate that for the quenches considered here, the ``thermal'' wavelength is comparable with the Compton wavelength.
\end{thebibliography}
\end{document}